\def\deg{\ensuremath{^\circ}}

\documentclass[
    ,final            % use final for the camera ready runs
  ]
  {aipproc}
\usepackage{epsfig}

\layoutstyle{8x11single}

\begin{document}

\title{Soft gamma-ray Galactic ridge emission as unveiled by SPI aboard INTEGRAL}


\author{J. Kn\"odlseder}{
  address={CESR (UPS/CNRS), 9, avenue du Colonel Roche, BP 44346, 
  31028 Toulouse Cedex 5, France}
}

\author{V. Lonjou}{
  address={CENBG, Universit\'e Bordeaux 1 (CNRS/IN2P3), Chemin du 
  Solarium, BP 120, 33175 Gradignan, France}
}

\author{G. Weidenspointner}{
  address={CESR (UPS/CNRS), 9, avenue du Colonel Roche, BP 44346, 
  31028 Toulouse Cedex 5, France}
}

\author{P. Jean}{
  address={CESR (UPS/CNRS), 9, avenue du Colonel Roche, BP 44346, 
  31028 Toulouse Cedex 5, France}
}

\author{A. Strong}{
  address={Max-Planck-Institut f\"ur extraterrestrische Physik, 
  Giessenbachstra§e, 85748 Garching, Germany}
}

\author{R. Diehl}{
  address={Max-Planck-Institut f\"ur extraterrestrische Physik, 
  Giessenbachstra§e, 85748 Garching, Germany}
}

\author{B. Cordier}{
  address={CEA Saclay, DSM/DAPNIA/Service d'Astrophysique, 91191 
  Gif-sur-Yvette, France}
}

\author{S. Schanne}{
  address={CEA Saclay, DSM/DAPNIA/Service d'Astrophysique, 91191 
  Gif-sur-Yvette, France}
}

\author{C. Winkler}{
  address={ESA/ESTEC, SCI-SA Astrophysics Division, 
  2201 AZ Noordwijk, The Netherlands}
}

\begin{abstract}
The origin of the soft gamma-ray (200 keV -- 1 MeV) Galactic ridge 
emission is one of the long-standing mysteries in the field of 
high-energy astrophysics. Population studies at lower energies
have shown that emission from accreting compact objects gradually 
recedes in this domain, leaving place to another source of gamma-ray 
emission that is characterised by a hard power-law spectrum extending 
from ~100 keV up to ~100 MeV. The nature of this hard component has 
remained so far elusive, partly due to the lack of sufficiently 
sensitive imaging telescopes that would be able to unveil the spatial 
distribution of the emission.
The SPI telescope aboard INTEGRAL allows now for the first time the
simultaneous imaging of diffuse and point-like emission in the soft
gamma-ray regime. We present here all-sky images of the soft gamma-ray
continuum emission that clearly reveal the morphology of the different
emission components. We discuss the implications of our results on the
nature of underlying emission processes and we put our results in perspective
of GLAST studies of diffuse Galactic continuum emission.
\end{abstract}

\maketitle

%%%%%%%%%%%%%%%%%%%%%%%%%%%%%%%%%%%%%%%%%%%%%%%%%%%%%%%%%%%%%%%%%%%%%%%%%%%%%%%%
% Introduction
%%%%%%%%%%%%%%%%%%%%%%%%%%%%%%%%%%%%%%%%%%%%%%%%%%%%%%%%%%%%%%%%%%%%%%%%%%%%%%%%
\section{Introduction}

The lack of sensitive imaging telescopes covering the soft gamma-ray 
(200 keV - 1 MeV) band has so far prevented the study of Galactic 
emission processes in this energy band.
Surveys at lower energies, such as those performed by HEAO-1 and RXTE, 
indicate that the Galaxy is dominated by strong point-source emission 
from X-ray binary systems.
In addition, a diffuse emission component is observed below $\sim60$ 
keV that may 
eventually be attributed to unresolved emission from a large number of 
CV systems \cite{krivonos07}.
Surveys at higher energies, such as those performed by the COMPTEL and 
EGRET telescopes aboard CGRO, reveal primarily diffuse emission from 
the Galactic plane.
While above $\sim100$~MeV the emission is relatively well understood 
as the result of cosmic-ray interactions with the interstellar 
medium, the observed 1-100 MeV emission is in excess of predictions 
from cosmic-ray propagation models \cite{strong00}.

The transition region around 200 keV - 1 MeV, where diffuse emission 
components should start to dominate point-source emission, has so far 
never been imaged.
The imaging spectrometer SPI aboard INTEGRAL combines for the first time 
good sensitivity and imaging performances to allow mapping the sky in 
this energy domain.
First attempts to detect the diffuse emission component with 
SPI using model fitting techniques have been reported by 
\cite{strong05} and \cite{bouchet05}.
The results clearly indicate the presence of a diffuse emission 
component that starts to dominate over the point-source emission 
above 100-200 keV.

In this paper we report about work in progress that aims in direct 
imaging of this diffuse emission.
The imaging is done using a multiresolution deconvolution procedure 
called MREM \cite{knoedl99,knoedl07}.
The data that were analysed in this work consist of those included 
in the December 8, 2006 public INTEGRAL data release.
They cover the period IJD 1074 - 2120, spanning almost 3 years of 
INTEGRAL mission data.
With this dataset, 85\% of the sky is covered with an exposure above 
10 ks, corresponding to a INTEGRAL/SPI point-source sensitivity of 
$\sim160$ mCrab in the $150-300$ keV band.
Most of the Galactic plane has an exposure in excess of 1 Ms, 
corresponding to a INTEGRAL/SPI point-source sensitivity of 
$\sim16$ mCrab in the $150-300$ keV band.

%%%%%%%%%%%%%%%%%%%%%%%%%%%%%%%%%%%%%%%%%%%%%%%%%%%%%%%%%%%%%%%%%%%%%%%%%%%%%%%%
% Skymaps
%%%%%%%%%%%%%%%%%%%%%%%%%%%%%%%%%%%%%%%%%%%%%%%%%%%%%%%%%%%%%%%%%%%%%%%%%%%%%%%%
\section{Skymaps}

\begin{figure}[!h]
  \includegraphics[width=.45\textwidth]{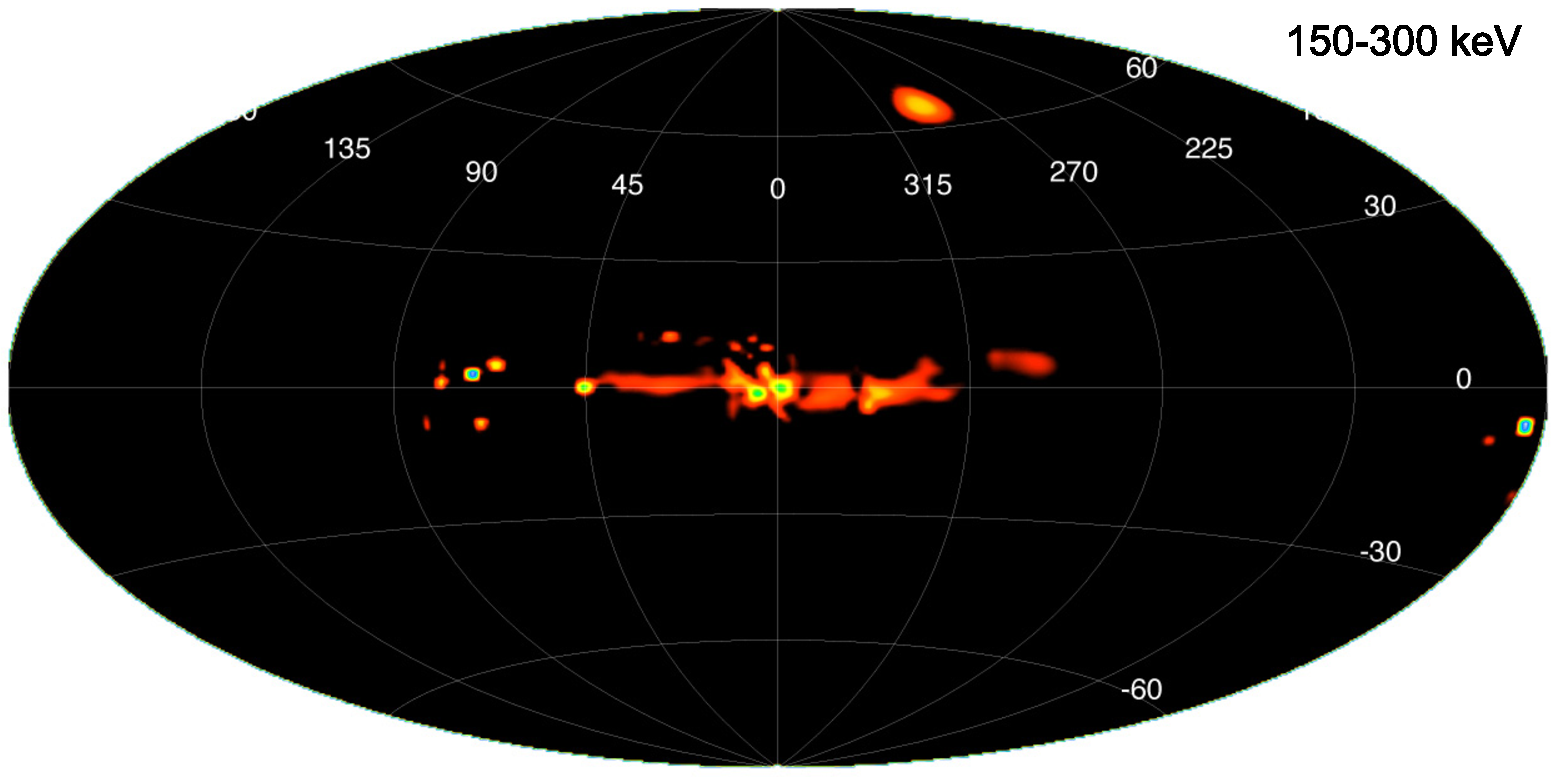}
  \hfill
  \includegraphics[width=.45\textwidth]{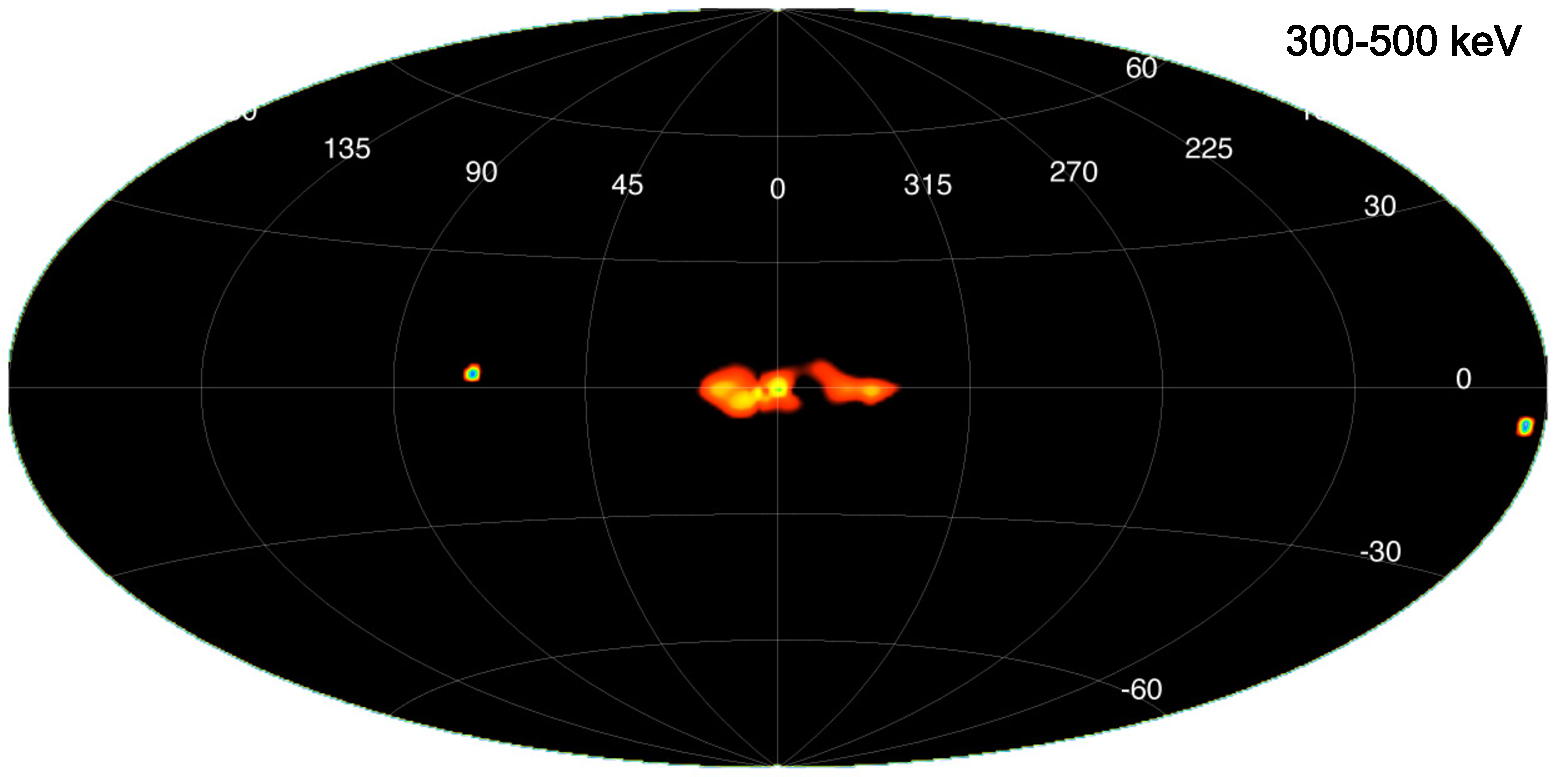}
  \caption{INTEGRAL/SPI multiresolution maps in the energy bands 
  150-300 keV (left) and 300-500 keV (right).}
  \label{fig:raw}
\end{figure}

Skymaps have been generated from INTEGRAL/SPI data using the MREM 
multiresolution deconvolution algorithm \cite{knoedl99,knoedl07} for 
a number of energy bands.
Due to the space limitations we show in Fig.~\ref{fig:raw} only the 
maps obtained for the 150-300 keV and 300-500 keV energy bands.
A full set of maps can be seen at 
\url{http://integral.esac.esa.int/POMNov2006.html}.

%%% SPI raw skymaps %%%%%%%%%%%%%%%%%%%%%%%%%%%%%%%%%%%%%%%%%%%%%%%%%%%%%%%%%%%%
%%%%%%%%%%%%%%%%%%%%%%%%%%%%%%%%%%%%%%%%%%%%%%%%%%%%%%%%%%%%%%%%%%%%%%%%%%%%%%%%

Figure \ref{fig:raw} clearly reveals the presence of diffuse emission 
in both energy bands. While in the 150-300 keV band still a number of 
point-sources are seen, they gradually fade away at higher energies.
Note that diffuse emission in this context does not necessarily mean 
that the underlying emission mechanism is indeed diffuse.
The apparent morphology may also be explained by a number of faint 
and unresolved point sources.

To reveal better the diffuse Galactic ridge emission, we remove known 
point-sources from the image by fitting their contribution in the 
data-space while deconvolving the residual counts into an allsky image.
The flux of each source in the catalogue is adjusted together with the 
level of the instrumental background during the image reconstruction 
procedure.
The multiresolution algorithm picks up any residual structure in the 
data that is not fit by the model components.
Some cross talk exists between diffuse emission and point-source 
emission due to the limited angular resolution of SPI 
(FWHM of $\sim3\deg$)
and the low detection significance for many of the sources.
Also numerous faint sources that are below the SPI detection threshold 
can not be resolved by the method.
Consequently, the residual emission seen in these {\em point-source 
subtracted skymaps} may be either explained by truly diffuse emission, 
the combined emission of weak and unresolved point sources, or by 
residual point source emission.

%%% SPI skymaps %%%%%%%%%%%%%%%%%%%%%%%%%%%%%%%%%%%%%%%%%%%%%%%%%%%%%%%%%%%%%%%%
\begin{figure}[!h]
  \includegraphics[width=.45\textwidth]{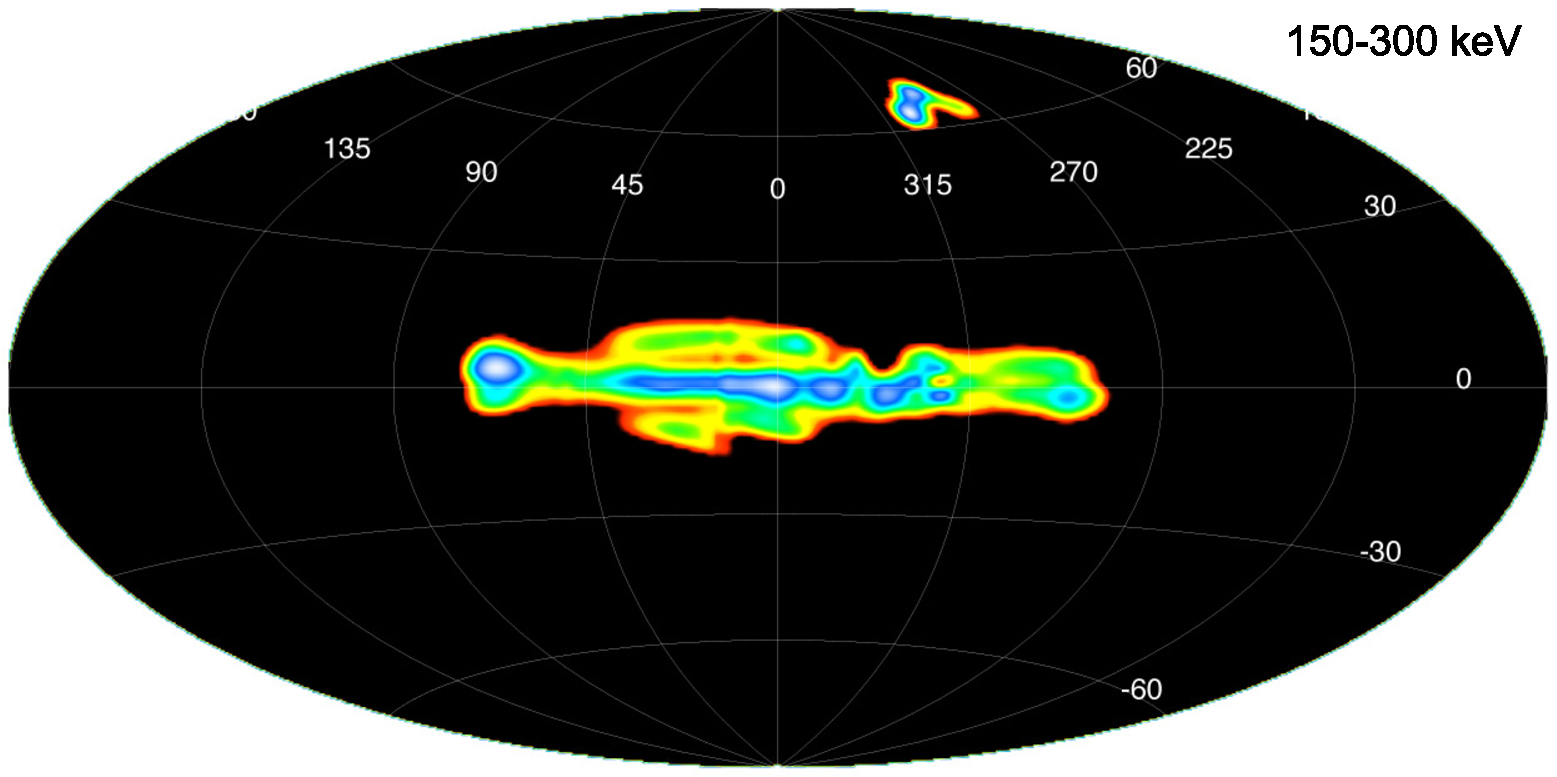}
  \hfill
  \includegraphics[width=.45\textwidth]{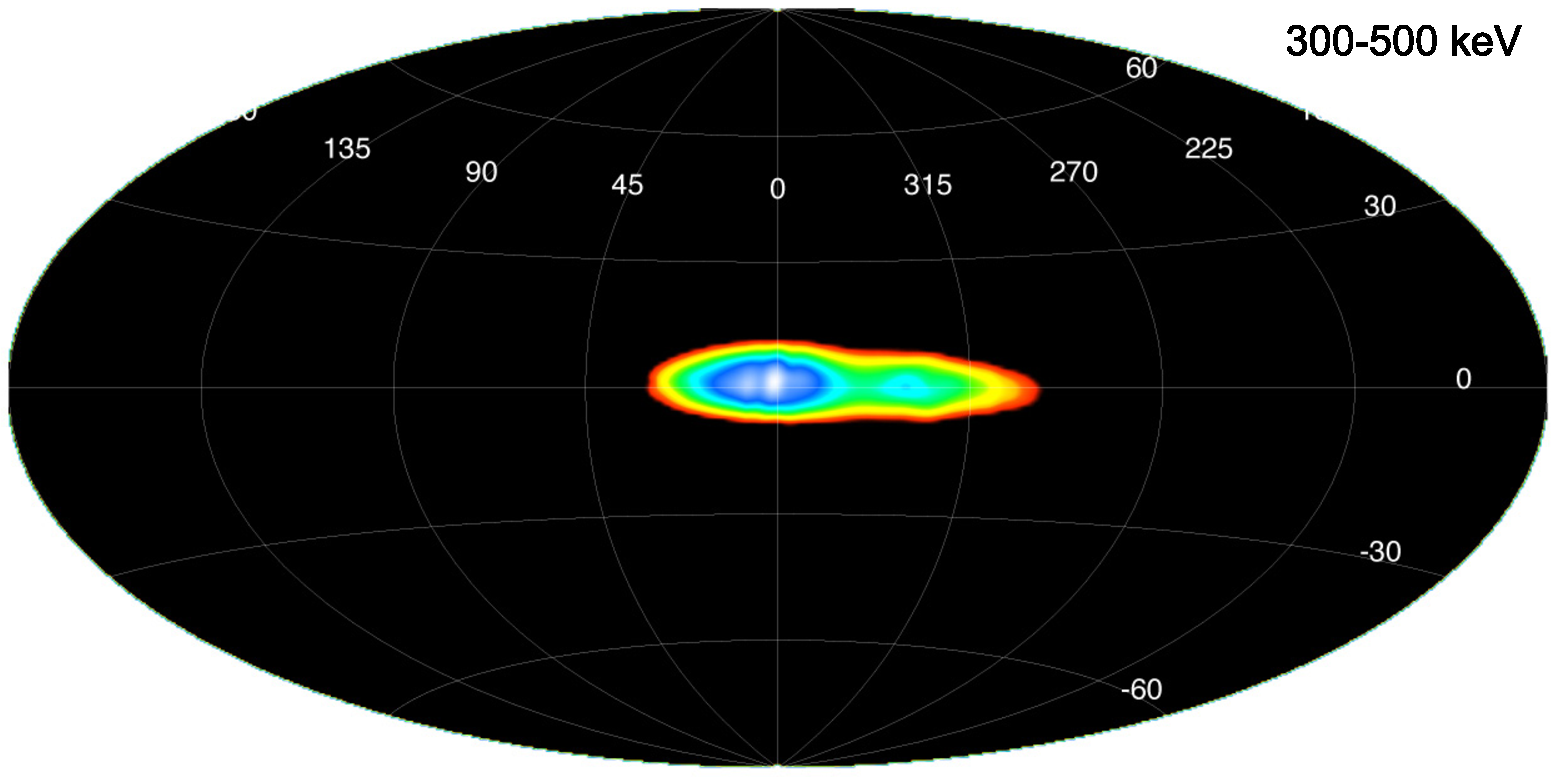}
  \caption{Point-source subtracted INTEGRAL/SPI multiresolution maps 
  in the energy bands 150-300 keV (left) and 300-500 keV (right).}
  \label{fig:res}
\end{figure}
%%%%%%%%%%%%%%%%%%%%%%%%%%%%%%%%%%%%%%%%%%%%%%%%%%%%%%%%%%%%%%%%%%%%%%%%%%%%%%%%

Figure \ref{fig:res} shows the point-source subtracted maps for the energy 
bands 150-300 keV and 300-500 keV.
In the 150-300 keV band an extended and narrow Galactic ridge emission is 
seen, reaching from the Cygnus region (at longitude $\sim80\deg$) to 
the Carina region (at longitude $\sim280\deg$).
Some of the structure seen in the map may be related to source 
variability, that has explicitly only be taken into account for the 
strongest known point-sources sources.
We therefore caution not to over interpret morphology details, and we 
are currently working on an improved treatment of source variability 
in the analysis.

In the 300-500 keV band the emission appears much more concentrated 
towards the inner Galaxy.
This concentration can be understood from the increasing contribution 
of Positronium continuum emission which amounts to roughly half of the 
Galactic emission seen in the image.
As we know from dedicated analyses, the positronium continuum 
morphology can be well described by a symmetrical Gaussian shaped 
bulge with FWHM of $\sim8\deg$ located at the Galactic centre 
\cite{weidenspointner06}.
Whether the apparent asymmetry of the underlying Galactic ridge 
emission is linked to positronium continuum emission or Galactic 
continuum emission remains to be examined by a dedicated spectral 
analysis.

%%%%%%%%%%%%%%%%%%%%%%%%%%%%%%%%%%%%%%%%%%%%%%%%%%%%%%%%%%%%%%%%%%%%%%%%%%%%%%%%
% Conclusions
%%%%%%%%%%%%%%%%%%%%%%%%%%%%%%%%%%%%%%%%%%%%%%%%%%%%%%%%%%%%%%%%%%%%%%%%%%%%%%%%
\section{Conclusions}

We have presented allsky maps of continuum emission in the 
150-300 keV and 300-500 keV energy bands that clearly show the 
presence of a diffuse or unresolved Galactic emission component.
Known diffuse emission mechanism, such as inverse Compton and 
bremsstrahlung processes, seem insufficient to explain the observed 
intensity levels (Strong et al., these proceedings).
In-situ electron acceleration could be a possible way to produce the 
observed emission without violating any energetic constraints 
\cite{dogiel02}.
In-flight annihilation of high-energy positrons or cosmic-ray 
knock-on secondary electrons could also contribute to the emission.

Alternatively, the apparently diffuse emission could arise from an 
unresolved population of weak sources.
Population synthesis indicates that a source population with a hard 
spectral index could indeed produce the observed emission without 
violating any source counts constraint \cite{strong06}.
If such a source class indeed exists it may also have implications on 
diffuse emission studies in the GLAST energy range.
Conversely, GLAST may even help to unveil the source population.
If, on the other hand, the soft gamma-ray Galactic ridge emission 
turns out to be intrinsically diffuse, the required modification of the 
cosmic-ray propagation models will necessarily also impact the 
analysis in the GLAST domain.
Building a coherent image of Galactic ridge high-energy emission will 
therefore be a major task within the next decade.

%%%%%%%%%%%%%%%%%%%%%%%%%%%%%%%%%%%%%%%%%%%%%%%%%%%%%%%%%%%%%%%%%%%%%%%%%%%%%%%%
% Acknowledgements
%%%%%%%%%%%%%%%%%%%%%%%%%%%%%%%%%%%%%%%%%%%%%%%%%%%%%%%%%%%%%%%%%%%%%%%%%%%%%%%%
\begin{theacknowledgments}
The SPI project has been completed under the responsibility and leadership 
of CNES.
We are grateful to ASI, CEA, CNES, DLR, ESA, INTA, NASA and OSTC for 
support.
\end{theacknowledgments}

\bibliographystyle{aipproc}   % if natbib is available

\end{document}